# Industrially Scalable Process for Silicon Nanowires for Seebeck Generation

G. F. Cerofolini[#1], M. Ferri[*2], E. Romano[#3], A. Roncaglia[*4], E. Selezneva[#5], A. Arcari[#6],
F. Suriano[*7], G. P. Veronese[*8], S. Solmi[*9], and D. Narducci[#10]

[#]Department of Materials Science, University of Milano-Bicocca, via R. Cozzi 53, 20125 Milano, Italy
[1]gianfranco.cerofolini@mater.unimib.it
[3]elisabetta.romano@mater.unimib.it
[5]ekaterina.selezneva@mater.unimib.it
[6]andrea.arcari1@unimib.it
[10]dario.narducci@unimib.it

[*]CNR-IMM, Via Gobetti 101, 40100 Bologna, Italy
[2]ferri@bo.imm.cnr.it
[4]roncaglia@bo.imm.cnr.it
[7]suriano@bo.imm.cnr.it
[8]veronese@bo.imm.cnr.it
[9]solmi@bo.imm.cnr.it

*Abstract*—The observation that the thermal conductivity of single-crystalline silicon nanowires with diameter on the length scale of 25 nm is lower than that of bulk material by two orders of magnitude has attracted the interest onto silicon as a potentially effective thermoelectric material. However, the potential interest has a hope of transforming in a practical interest only if poly-crystalline silicon can replace single crystalline silicon and the preparation of nanowires does not involve any advanced photolithography. In this work we show that a technique, based on the controlled etching and filling of recessed regions and employing standard photolithography and deposition-etching methods, succeeds in the preparation of poly-crystalline silicon nanowires (with diameter of 25 nm and length on the centimetre scale) at a linear density of $3 \times 10^6$ cm$^{-1}$.

I. INTRODUCTION

Global warming [1], the nearby (or already surpassed) Hubbert's oil peak [2], the limited amount of hydrocarbon reserves [3], and the appearance on the oil market of large consumers like China and India, have fuelled the interest towards the exploitation of renewable or waste energies. With 'renewable energy' we intend any form of energy distributed more or less homogeneously on the Earth surface and whose primary sources are expected to exhaust on a time scale longer than the historic one. With 'waste energy' we intend all forms of energy which is unavoidably produced in the transformation of chemical energy into mechanical or electrical energy and dissipated as low-temperature enthalpy. Approximately 30% of the whole energy production is lost as waste energy.

In view of the Second Law of Thermodynamics, the exploitation of waste energy is necessarily a low-yield process. However, while the exploitation of of renewable energy generally involves large areas (e.g., geothermal or photovoltaic fields), the exploitation of waste energy can be based on devices integrated in existing apparatuses. In a way, devices for the exploitation of waste energy can be thought of as relatively small and their fabrication might involve techniques taken from fields like microelectronics, although materials and processes must be very cheap because of the inherently low thermodynamic yield .

Among the possible routes for the exploitation of waste energy, thermoelectricity is gaining an increasing interest. The thermoelectric properties of any material are generally described by a quantity $Z$ including the electrical resistivity $\rho$, thermal conductivity $\kappa$, and thermoelectric coefficient $\alpha$ of the material:

$$Z = \alpha^2/\rho\kappa.$$

The reason for the importance of such a quantity is that the maximum thermodynamic yield $\eta$ of a thermoelectric generator operating between temperatures $T_1$ and $T_2$ ($T_1 < T_2$) is negligible for $ZT \ll 1$, while is close to that of the ideal Carnot engine for $ZT \gg 1$, where $T$ is an intermediate temperature between $T_1$ and $T_2$, say $T = \frac{1}{2}(T_1 + T_2)$. A thermoelectric material is commonly assumed to be of potential interest for practical applications when $ZT \gtrsim 1$.

Bulk single-crystalline silicon is notoriously a material of poor interest for its thermoelectric properties. Combining its bulk thermal conductivity ($\kappa = 1.3$ W cm$^{-1}$K$^{-1}$ at 300 K [4] vs. that of a good thermal insulator like SiO$_2$, $\kappa = 0.01$ Wcm$^{-1}$K$^{-1}$), its electrical resistivity ($\rho \approx 1$ mΩcm for degenerate materials vs. $0.1 - 1$ μΩ cm characteristic of metals), and its thermoelectric power (ranging from

$|\alpha| \simeq 0.5$ µV K$^{-1}$ at carrier concentration of $10^{18}$ cm$^{-3}$ to $|\alpha| \simeq 0.2$ µV K$^{-1}$ at $10^{20}$ cm$^{-3}$ in both p- and n-type at 300 K) gives $ZT \approx 0.01$ at room temperature.

Although silicon is a relatively inexpensive material and we have a fantastic technology for its processing, its figure of merit is so low compared with that of a good thermoelectric material (like Bi$_2$Te$_3$, for which $ZT = 0.9$ at room temperature) as to discard it as an interesting material for thermoelectric generation.

This situation has however significantly changed with the observation that thermal diffusion in silicon wires decreases significantly when the wire diameter is reduced to the 10-nm length scale (by a factor of ca. 50 ongoing from bulk silicon to 22-nm nanowires, for which $\kappa \simeq 6 \times 10^{-2}$ W cm$^{-1}$K$^{-1}$ at 300 K [5]) and that their surface can be finished (ruggidizing it in a suitable roughness regime) to reduce further the thermal conductivity [6]. This dramatic decrease of thermal conductivity is attributed to a combination of the essentially one-dimensional nanowire (NW) geometry with the ability of rugged surfaces to scatter phonons [6], [7]. Since in heavily doped silicon the electron mean free path is appreciably smaller that 20 nm, the reduced NW size should not affect appreciably its resistivity, so that silicon nanowire should have a thermoelectric figure of merit not so far from unity ($ZT \approx 0.35$ at 300 K and $10^{20}$ cm$^{-3}$ for NW diameter of 10 nm [7]) thus close to values of interest for applications.

The above conclusions were taken studying silicon NWs prepared from single-crystalline silicon and employing methods that cannot be scaled for large volume applications. In particular, in [6] silicon NWs were prepared by a complex process involving the oxidation of single-crystalline silicon with Ag$^+$, the dissolution in HF$_{aq}$ of the oxide so formed, the removal and cleaning of the remaining NWs. In [7], instead, thinned silicon-on-insulator was used as substrate and imprint lithography (using masks prepared by the superlattice nanowire pattern transfer technique) was employed for the definition of the geometry.

Needless to say, the use of Si NWs for large-scale Seebeck generation passes through the use of polycrystalline silicon (poly-Si) as basic material and the development of an adequate technology for the large-scale production of NWs.

Although poly-Si has, for assigned doping level, a somewhat higher resistivity than that of single crystalline silicon, its thermal conductivity is lower (typically by $1-2$ orders of magnitude [8]) and grain boundaries may perhaps be beneficial on $|\alpha|$ because of their effect on energetic filtering, as proposed in [9]. In view of this combination of factors, the thermoelectric performance of poly-Si NWs is expectedly competitive, if not better, with that of single crystalline silicon. In this work we accept, without further discussion although tentatively, this *Ansatz* and concentrate our attention on the development of a cost-effective process for the preparation of poly-Si NWs.

## II. THE BASIC IDEA

The first obvious choice to reduce production costs involves the deposition of poly-Si, whose industrial cost is essentially determined by its preparation technique (pyrolytic decomposition of a precursor, silane in our case), onto relatively inexpensive insulating substrates.

The cost of its production in NWs is essentially related to the NW production technology. The techniques developed for the production of NW arrays without using advanced lithography (nanoimprint lithography, NIL [10], [11], [12] or multisidewall patterning techniques, MSPTs [14], [15], [16], [17], [18]) suffer from a fundamental difficulty: even assuming they can attain the theoretical packing limit for a planar arrangement, they succeed in the production of a single layer of wires arranged along the surface.

The maximum linear density achievable with any of the considered techniques is controlled by the NW diameter: assuming this diameter of about 25 nm (not to have appreciable reduction of carrier mobility) and a pitch twice the diameter, the maximum achievable NW density would be of $2 \times 10^5$ cm$^{-2}$. Increasing the density is manifestly possible defining a stack of several NW arrays, each running parallel to the surface. However, when using NIL or MSPTs this procedure requires one lithographic step per layer of the stack at least, and thus seems poorly consistent with the need of cost effectiveness of thermoelectricity. It follows that to be of potential interest in practical applications *any process for Seebeck nanowires must be able to prepare,* without the reiteration of the process, *arrays not only with close packing in the plane but also arranged vertically.*

This task is possible via a method, developed by the present authors, involving the deposition of a suitable multilayered stack and the *controlled etching and filling of recessed regions* (CEFRR) resulting from its processing [19], [20], [21].

The basic idea underlying the CEFRR technique is sketched in Fig. 1:

(a) The process starts with the deposition on an electrical and thermal insulator ('substrate') of a stack of $N$ insulating bilayers

$$\overbrace{\overbrace{\underbrace{A|B|}_{1}}^{t^A+t^B}\overbrace{\underbrace{A|B|}_{2}}^{t^A+t^B}\cdots\overbrace{\underbrace{A|B|}_{N}}^{t^A+t^B}}^{N(t^A+t^B)}, \quad (1)$$

where $t^A$ and $t^B$ are the thicknesses of the constituting layers and the total thickness $t_N$ is given by $t_N = N(t^A + t^B)$. The intended example of bilayer is SiO$_2$|Si$_3$N$_4$.

(b) The multilayered stack is patterned via highly directional attack (reactive ion etching, RIE) with the formation of deep trenches stopping with the exposure of the substrate.

(c) The exposed walls are selectively etched with the formation of recessed regions whose extension depends on the duration of etching (in the intended example the layer undergoing the selective etching with HF$_{aq}$ is SiO$_2$).

(d) The recessed regions are filled via conformal chemical vapour deposition (CVD) at low pressure of poly-Si.

(e) The poly-Si undergoes controlled oxidation to an amount sufficient to form separate nanowires. An etching of the

newly formed $SiO_2$ may be done if required to contact the wire.

Although in the intended example the bilayer is $SiO_2|Si_3N_4$, A and B may actually be any two good *thermal and electrical insulators*, characterized by the existence of a selective etching for A with respect to B.

If the process resulted in truly vertical walls, the NW density would limited by the number $N$ of bilayers forming the stack. However, the directionality of RIE is not absolute so that the trench will be characterized by a finite aspect ratio $R$ (see Fig. 2).

As discussed in [22], the finite value of $R$ limits the achievable density to a supremum $\delta_{\text{sup}}$ given by

$$\delta_{\text{sup}} = \frac{R}{t^A + t^B}. \qquad (2)$$

The vertical arrangement allows thus a magnification by a factor of $R$ of the maximum density achievable for a planar arrangement of closely packed nanowires. Assuming $t^A + t^B = 50$ nm and $R = 20$ (an aggressive, but not exaggerated, value), Eq. (2) predicts a *limiting linear density, of $4 \times 10^6$ cm$^{-1}$, larger by a factor of 7 than the most dense structure hitherto presented* [23]. The limiting value is approached when $N$ exceeds a characteristic value $N^*$,

$$N \gg N^* \Longrightarrow \delta \simeq \delta_{\text{sup}}, \qquad (3)$$

with

$$N^* = \frac{(W+s)R}{2(t^A + t^B)}. \qquad (4)$$

where $W$ is the stack width at its top and $s$ is the separation of neighbouring stacks at their bases.

Equation (2) says that the limiting density does not depend on lithography; Eqs. (3) and (4) state however that the lithography impacts on the technology, determining the value of $N^*$. For $R = 20$ and $t^A + t^B = 50$ nm, a (relaxed) lithography with $W + s = 2$ μm would give $N^* = 400$ and consequently a stack with height $N(t_A + t_B)$ taller than 20 μm, whereas an (aggressive) technology with $W + s = 0.2$ μm would give $N^* = 40$ and consequently a stack taller than 2 μm, much easier to manage. In the latter case the need of cost effectiveness would however require a sublithographic definition of the stack.

### III. Proving the idea

The CEFRR technique
(i) requires only one non-critical lithographic definition (say on the 1-μm length scale),
(ii) involves only few fabrication steps, well known in silicon device processing,
(iii) may be carried out on insulating substrates with no special specifications (in terms of flatness, purity, crystallinity, etc.), and
(iv) the initial mask may be designed to allow the production of NW arrays with length (in the range $10^3 - 10^4$ μm) adequate for macroscopic handling and with linear density of the order of magnitude of $10^6$ cm$^{-1}$.

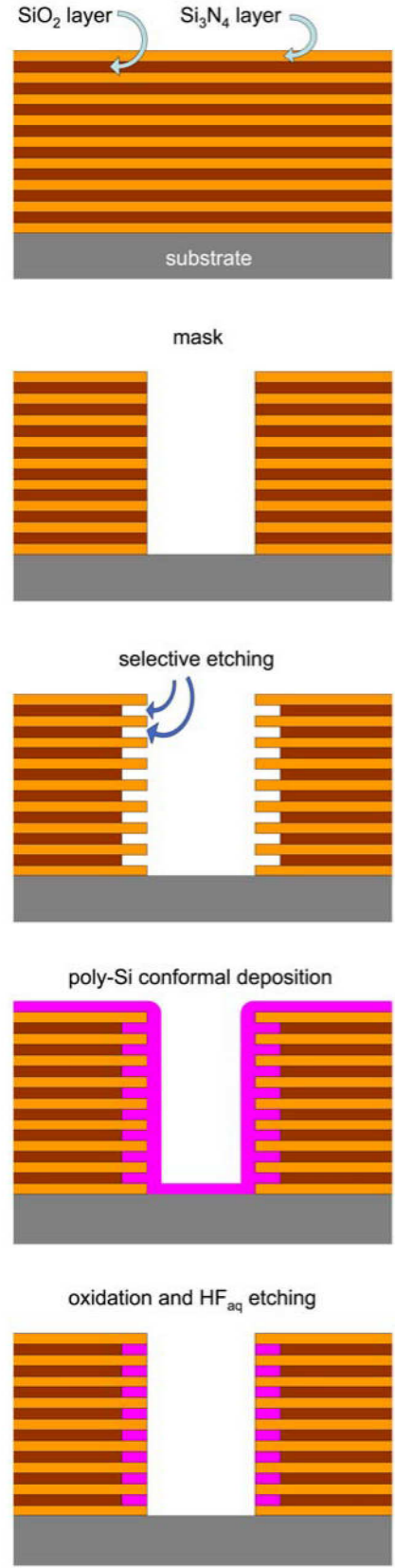

Figure 1. Preparation of a uniform array of nanowires arranged perpendicularly the surface

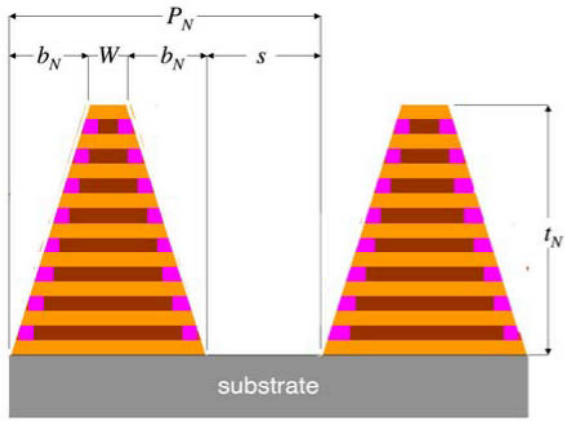

Figure 2. Geometric parameters defining the trench

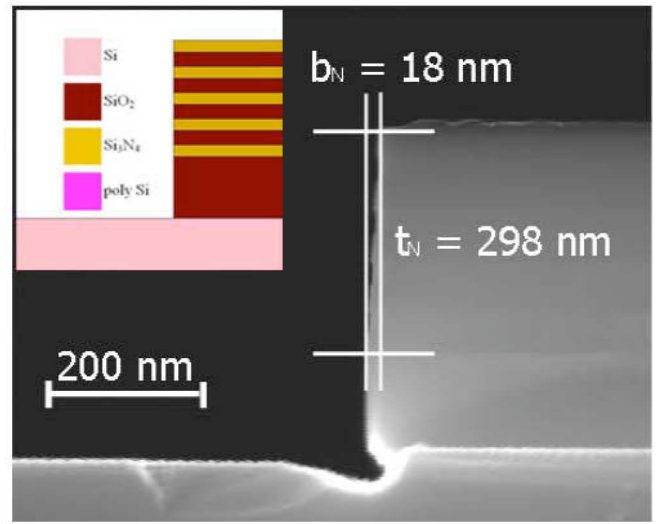

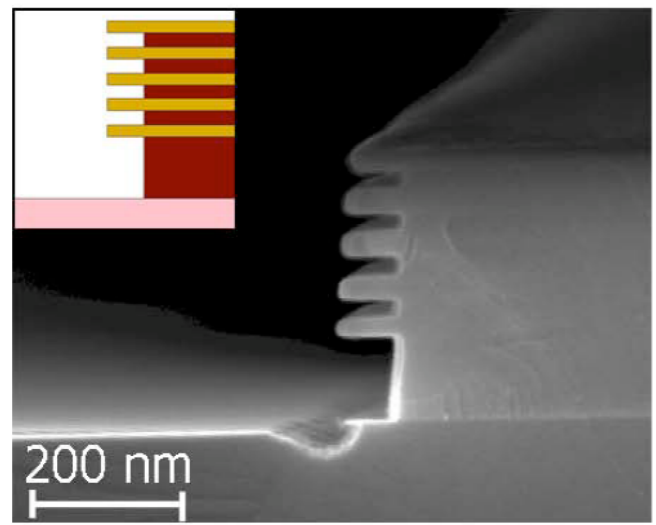

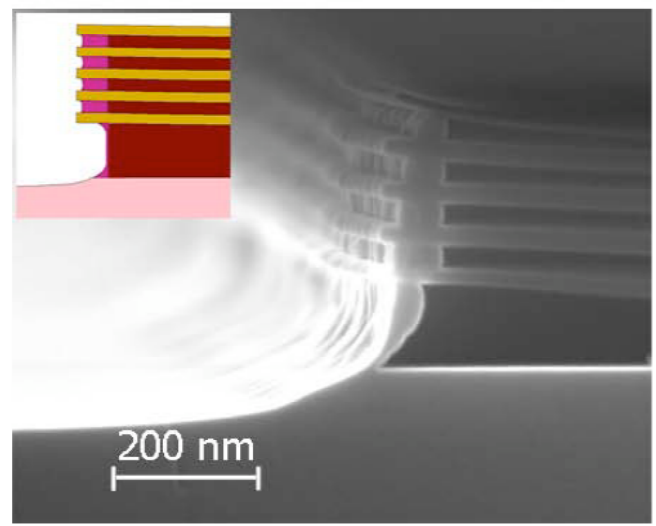

Figure 3. SEM images of the structures resulting after step (b), reactive ion etch (*top*); step (c), formation of the recessed regions via $HF_{aq}$ etching (*middle*), and step (e), formation of the NWs (*bottom*)

The combination of all these features makes the CEFRR technique an ideal candidate for the preparation of silicon-based Seebeck generators.

To prove the idea we prepared according to process (a)–(e) a stack of 4 bilayers. To decrease the thermal conductivity without increasing the electrical resistivity the NW diameter must be shorter than the phonon mean free path but much larger than the electron mean free path. Since the electron mean free path decreases monotonically with the carrier concentration one could be tempted to reduce as far as possible the NW diameter counterbalancing this effect via an increase of carrier concentration. This operation, however is contrasted by two factors: the progressive loss of the bulk-silicon properties when $d$ approaches the nanometre length scale and the decrease of $|\alpha|$ with dopant concentration predicted by the elementary theory of the Seebeck effect. Keeping the $SiO_2$ thickness to 30 nm, the structures resulting after steps (b), (c) and (e) were imaged by scanning electron microscopy (SEM) and are shown in Fig. 3.

Although the process was developed in an academic facility with 2-μm lithography, Fig. 4 shows that the process can be scaled down to a line width of approximately 0.2 μm at least (result achieved using the sidewall patterning technique [15], [17] for its patterning) without any appreciable degradation.

Of course, the SEM imaging do not provide adequate information on electrical continuity of the NWs. The integrity of the NWs was thus tested by measuring their electrical conductance. To obtain highly conductive NWs the poly-Si was doped by a conventional process formed by a predeposition in $POCl_3$ at 920 °C followed by an annealing in an inert ($N_2$) atmosphere at 1100 °C.

Providing the NWs with ohmic contacts using standard Al:Si metallization was found easier than expected. We found that the conductance of the parallel of $n$ NWs ($n = 8, 16, 32$) increases in proportion to $n$ and decreases with the NW length $L$ as $L^{-1}$ (for $L$ in the interval $1 - 4$ μm). Two- and four-pont measurements showed differences in all cases smaller than 0.2%, thus indicating that the contact resistance does not affect appreciably the conclusion. A few measurements on NWs with length on the 10-cm length scale confirmed the electrical continuity of the wires.

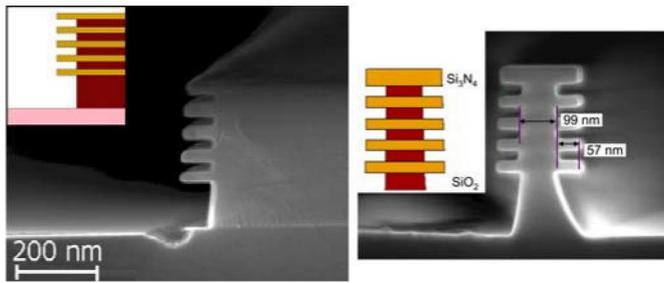

Figure 4. SEM image of the cross section of the recessed regions formed at the side of a photolithographic (*left*) or a sublithographically defined seed (*right*)

## IV. Conclusions

A cost effective method for the production of poly-Si NWs has been developed. This method requires standard lithography and employs familiar processes of the silicon technology. It succeeds in the preparation of nanowires with diameter on the length scale of 30 nm, contactable with negligible contact resistance using standard Al:Si metallization, and preserving electrical continuity on macroscopic length (proved up to 10 cm). The maximum achievable linear density depends on the employed techniques, but is anyway larger by one order than those achievable by NIL or MSPTs.